\documentclass[a4paper,11pt]{article} 
  
\usepackage[english]{babel}
\usepackage[a4paper]{geometry}
\usepackage{amsmath}
\usepackage{amsthm}
\usepackage{amssymb}
\usepackage{color}

\usepackage{hyperref}         

\def\bR{\mathbb{R}}

\def\bN{\mathbb{N}}

\def\bZ{\mathbb{Z}}
\def\cC{\mathcal{C}}

\def\cQ{\mathcal{Q}}

\def\cM{\mathcal{M}}
\def\cR{\mathcal{R}}

\def\cV{\mathcal{V}}
\def\cO{\mathcal{O}}
\def\cF{\mathcal{F}}
\def\cG{\mathcal{G}}
\def\cL{\mathcal{L}}
\def\cJ{\mathcal{J}}
\def\cN{\mathcal{N}}
\def\cE{\mathcal{E}}
\def\cK{\mathcal{K}}
\def\cH{\mathcal{H}}

\def\ph{\varphi}

\def\indic{\hbox{\raise-2pt \hbox{\indbf 1}}}

\let\dpr=\partial

\let\==\equiv

\let\io=\infty
\let\0=\noindent

\def\*{{\hfill\break\null\hfill\break}}

\def\bmedia#1{{\bigl\langle#1\bigr\rangle}}

\def\eg{\hbox{\it e.g.\ }}

\def\tende#1{\,\vtop{\ialign{##\crcr\rightarrowfill\crcr
             \noalign{\kern-1pt\nointerlineskip}
             \hskip3.pt${\scriptstyle #1}$\hskip3.pt\crcr}}\,}
\def\otto{\,{\kern-1.truept\leftarrow\kern-5.truept\to\kern-1.truept}\,}

\def\tr{{\rm tr}}

\newtheorem{theorem}{Theorem}[section]  

\newtheorem{prop}[theorem]{Proposition}
\newtheorem{lemma}[theorem]{Lemma}

\numberwithin{equation}{section}

\def\tl#1{{\tilde{#1}}}
\def\be{\begin{equation}}
\def\ee{\end{equation}}

\newcommand{\hc}{\mbox{h.c.}}

\let\a=\alpha \let\b=\beta    \let\g=\gamma     \let\d=\delta     
        \let\k=\kappa     \let\l=\lambda
\let\m=\mu                          \let\r=\rho
\let\s=\sigma \let\t=\tau         \let\ph=\varphi   
\let\ps=\psi        
 \let\D=\Delta       \let\L=\Lambda    
             
\let\O=\Omega

\usepackage{tikz}
\usetikzlibrary{fit}
\usetikzlibrary{shapes.geometric}
\usetikzlibrary{decorations.pathmorphing}

\tikzset{
point/.style={circle,fill=black,inner sep=1pt},
vertex/.style={circle,fill=black,inner sep=1.5pt},   
bvertex/.style={circle,fill=black,inner sep=2.8pt},
Bvertex/.style={circle,fill=black,inner sep=4pt}, 
specialEP/.style={rectangle,fill=white,draw,inner sep=3pt},  
whitevex/.style={circle,fill=white,draw, inner sep=2pt},
linelabel/.style={sloped,above,very near start, inner sep=1pt,execute at begin node=$\scriptstyle,execute at end node=$},
baseline=(current  bounding  box.center),doubled/.style={double distance= 1pt,line width=1.5pt},
th/.style={line width=0.5 pt, gray},  
med/.style={line width=1 pt}  
}

\begin{document}

\title{Bogoliubov theory for dilute bose gases: \\ the Gross-Pitaevskii Regime}

\author{Serena Cenatiempo \\ \\
Gran Sasso Science Institute, \\ Viale Francesco Crispi 7, 
67100 L'Aquila, Italy}

\maketitle

\begin{abstract}
 In 1947 Bogoliubov suggested a heuristic theory to compute the excitation spectrum of weakly interacting Bose gases.  Such a  theory predicts a linear excitation spectrum and provides expressions for the thermodynamic functions which are believed to be correct in the dilute limit. 
Thus far, there are only a few cases where the predictions of Bogoliubov can be obtained by means of rigorous mathematical analysis. A major challenge is  to control the corrections beyond Bogoliubov theory, namely to test the validity of  Bogoliubov's predictions in regimes where the approximations made by Bogoliubov are not valid. 
In these notes we discuss how this challenge can be addressed in the case of a system of $N$ interacting bosons trapped in a box with volume one in the Gross-Pitaevskii limit, where the scattering length of the potential is of the order $1/N$ and $N$ tends to infinity. 
This is a recent result obtained in \cite{BBCS3} and \cite{BBCS4}, joint works with C. Boccato, C. Brennecke, and B. Schlein, which extend a previous result obtained in \cite{BBCS1}, removing the assumption of small interaction potential.




\end{abstract}

\section{Introduction}

Since the early experiments on superfluidity in liquid helium \cite{He1, He2}, and even more after the first experimental realizations of Bose-Einstein condensation in cold atomic gases \cite{BEC1,BEC2,BEC3}, the understanding of the low temperature properties 
of systems of interacting bosons has stimulated several theoretical and mathematical investigations.
%
%
Aim of these notes is to report on a recent result establishing the equilibrium properties of the interacting Bose gas in one of the regimes which are relevant for the description of condensation 
in low-interacting and dilute atomic gases, the so called {\it Gross-Pitaevskii regime} which will presented below\footnote{We refer the reader to \cite{BPS}, and to the recent result \cite{BS}, for a summary of the results concerning the time evolution of Bose-Einstein condensates.}. As a prelude, we will start by reviewing the progresses made so far in the comprehension of the equilibrium properties of the interacting Bose gas in the thermodynamic limit. This preliminary discussion will set the stage to clarify the mathematical difficulties posed by the Gross-Pitaevskii regime, and to compare the main result obtained in \cite{BBCS3, BBCS4}  with related results  achieved in different parameters regimes.



In the course of these notes we are going to consider systems of $N$ bosons in a three dimensional box $\Lambda$ of size $L$  with periodic boundary conditions (in the more general case the bosons are rather trapped by an external confining potential). 
The Hamilton operator describing the system has the form 
\begin{equation}\label{eq:Ham0} 
H_{N,\L} = \sum_{j=1}^N -\Delta_{x_j} +  \sum_{i<j}^N V ( x_i -x_j) \end{equation}
and acts on the Hilbert space $L^2_s (\Lambda^N)$, the subspace of $L^2 (\Lambda^N)$ consisting of functions that are symmetric with respect to permutations of the $N$ particles. We require $V$ to be non-negative, radial, and to have finite zero energy scattering length~$\frak{a}_0 $. The latter is defined as 
\be \label{eq:fraka0}
\frak{a}_0 = (8 \pi)^{-1} \int V(x)f(x)\,,\ee with $f(x)$ solution of the zero energy scattering equation $(-\D +\frac 12 V(x)) f(x)=0$, with boundary condition $f(x)\to 1$ as $|x|\to \io$. We will first discuss the equilibrium properties of the system in the {\it thermodynamic limit}, where the density of the system $\r = N/|\L|$ is kept constant and the size of the box $\L$ is sent to infinity.
It is well known that in absence of interaction the systems exhibits Bose-Einstein condensation; in particular at zero temperature all particles are in the ground state of the kinetic energy operator, which is given by the zero momentum mode. A long-standing goal is to understand what happens to the system when we take into account the interaction among particles. Does the system still exhibit condensation? Can we provide expressions for the ground state energy and excitation spectrum, at least in some weakly interacting regime? Can we explain the emergence of super-fluidity, as observed in experiments? 

The usual picture of Bose-Einstein condensation in the homogeneous interacting case is based on an approximate exactly solvable model due to Bogoliubov \cite{B} (see also \cite[Appendix A]{LSSY} for a review). Bogoliubov rewrote the Hamilton operator (\ref{eq:Ham0}) in momentum space, using the formalism of second quantization. Since he expected low-energy states to exhibit Bose-Einstein condensation (at least for sufficiently weak interaction), he replaced all creation and annihilation operators associated with the zero-momentum mode  by factors  $N^{1/2}$. The resulting Hamiltonian contains constant terms (describing the interaction among particles in the condensate), terms that are quadratic in creation and annihilation operators associated with modes with momentum $p \not = 0$ (describing the kinetic energy of the excitations as well as the interaction between excitations and the condensate) and terms that are cubic and quartic (describing interactions among excitations). Neglecting all cubic and quartic contributions, Bogoliubov obtained a quadratic Hamiltonian that he could diagonalize explicitly, obtaining the following expression for the ground state energy
\be  \label{eq:BogEnergy} \begin{split}
E_{N,\L} =\; & \frac{N }{2} \rho \widehat V(0) - \tfrac 1 4 \hskip -0.1cm\sum_{\substack{p \in \frac{2\pi}{L} \bZ^3 \\ p \neq 0}} \frac{ (\rho \widehat V(p) )^2 }{p^2} \\
&- \tfrac 1 2 \sum_{\substack{p \in \frac{2\pi}{L} \bZ^3 \\ p \neq 0}} \Big[ p^2 + \rho \widehat V(p) - \sqrt{p^4  \hskip -0.05cm+  \hskip -0.05cm2 \rho \widehat V(p) p^2 } - \frac 1 4   \frac{ (\rho \widehat V(p) )^2 }{p^2} \Big] \,,
\end{split}\ee
and an  excitation spectrum 
of the form\footnote{The linearity for small momenta of the expression \eqref{eq:BogSpec} for the excitation spectrum was used by Bogoliubov to explain the emergence of superfluidity, via the so-called Landau criterion \cite{Lan}.}
\be  \label{eq:BogSpec}
\sum_{p \in \frac{2\pi}{L} \bZ^3} \sqrt{p^4 + 2\r  \widehat V(p)p^2}\, n_p 
\ee
for finitely many $n_p \in \bN$.
%
%
Remarkably, Bogoliubov recognized that, after having taken the thermodynamic limit, the expressions 
\[
\frak{a}^{(0)}_0= (8 \pi)^{-1} \widehat V(0)\,, \qquad  \frak{a}^{(1)}_0= -  \int \frac{d^3 p}{(2\pi)^3} \frac{\widehat V(p)^2}{2p^2}  
\]
appearing on the r.h.s. of  \eqref{eq:BogEnergy}, were just the first and second Born approximations of the infinite volume scattering length $\frak{a}_0 $. By replacing the sum $\frak{a}^{(0)}_0 + \frak{a}^{(1)}_0$  by  $\frak{a}_0 $ in the first line of Eq. \eqref{eq:BogEnergy}, and $\widehat V(p)$ by $8 \pi \frak{a}_0 $ in the second line of the same equation\footnote{It is easy to see that as $\r \to 0$ only very small $p$ will play a role in the integral on the second line of \eqref{eq:BogEnergy}, hence we can substitute $\widehat V(p)$ by $\widehat V(0)$ up to errors of smaller order than $ N \frak{a_0} (\frak{a_0}\r)^{3/2}$.},  Bogoliubov obtained the following formula 
%
for the ground state energy for particle of a dilute bose gas in the thermodynamic limit
\begin{equation}\label{eq:LHY}  \lim_{\substack{N,|\L| \to \infty \\ \rho = N / |\L|}} \frac{E_ {N,\L}}{N} = 4\pi \rho \frak{a}_0 \left[ 1 + \frac{128}{15 \sqrt{\pi}} (\rho \frak{a}_0^3)^{1/2} + o ((\rho \frak{a}_0^3)^{1/2} ) \right]\,, \end{equation}
which is known as Lee-Huang-Yang formula \cite{LeeYang, LHY}. A similar substitution is expected to give the correct expression for velocity of sound
\[
v_s = \lim_{p \to 0} \frac{\sqrt{p^4 + 16 \pi \r \frak{a}_0 p^2}}{p} = \sqrt{16 \pi \r \frak{a}_0}\,,
\]
obtained by substituting $\widehat V(p)$ by $8 \pi \frak{a}_0$ in the dispersion relation provided by \eqref{eq:BogSpec}. As an additional example one can compute within Bogoliubov approximation the expected density of particles outside the condensate in the ground state (the so called {\it condensate depletion}), obtaining
\begin{equation}\label{eq:depl0} 
\rho_+ =  \sum_{\substack{p \in \frac{2\pi}{L} \bZ^3 \\ p \neq 0}} \left[ \; \frac{p^2  + \r \widehat V(p)  - \sqrt{p^4 + 2 \r \widehat V(p)   p^2}}{2 \sqrt{p^4 + 2 \r \widehat V(p)  p^2}}  \; \right] \,. \end{equation}
Taking the thermodynamic limit of \eqref{eq:depl0} and  substituting $\widehat V(p)$ with $8 \pi \frak{a}_0$ one obtains the prediction
\be \label{eq:depl1} 
\frac{\rho_+}{\r}=  \frac{8}{3 \sqrt \pi}  \sqrt{\r \frak{a}_0^3}\,.
\ee



It is worth to stress that Bogoliubov 's model is based on the very strong assumption (not a priori justified) that the interacting system exhibits condensation, plus a quite rough truncation of the Hamiltonian. Nevertheless Bogoliubov predictions are believed to be correct in the dilute limit $\r \frak{a}_0^3 \ll 1$, the reason being that in his final replacement Bogoliubov might compensate exactly for all  terms (cubic and quartic in creation and annihilation operators) that he neglected in his analysis\footnote{Notice also that both the Lee-Huang-Yang formula \eqref{eq:LHY} and the prediction \eqref{eq:depl1} for the condensate depletion have been also recently measured in experiments \cite{ExpLHY, ExpDepletion}.}. It is then not surprising that Bogoliubov '47 paper was followed by several attempts to study in a systematic way the corrections to Bogoliubov theory, that is to understand the role of the cubic and quartic contributions neglected in Bogoliubov approximation.
 Unfortunately  perturbation theory around Bogoliubov model is plagued by ultraviolet and infrared divergences, whose meaning could be in principle that the interacting system has completely different physical properties with respect to the ones predicted by Bogoliubov. 
A few partial results confirming Bogoliubov picture have been obtained in the late '60s on the basis on diagrammatic techniques borrowed from Quantum Field Theory \cite{Beliaev, HugPines, LeeYang, LHY, GavNoz, NepNep, PopSer, Ben, PCCS1, PCCS2}, but they were all based on the summations of  special classes of diagrams selected from the divergent perturbation theory.

More recently, a study of the whole perturbation theory around Bogoliubov model for weak repulsive interactions (and/or at low densities) in three dimension, and the proof of its order by order convergence after proper resummations, has been obtained by Benfatto \cite{B}. This work provides a strong indication of the stability of three dimensional Bose-Einstein condensate at zero temperature, and a confirmation of the expression \eqref{eq:BogSpec} with a renormalized speed of sound\footnote{The method employed by Benfatto in \cite{B} is the Wilsonian Renormalization Group, combined with the ideas of constructive renormalization group, in the form developed by the roman school of Benfatto, Gallavotti et al since the late seventies.  See also \cite{PCCS1, PCCS2} for similar theoretical physics results obtained by means of dimensional regularization.}.
%
%
It is worth stressing that even though the method used by Benfatto is taken from the constructive theory, the resulting bounds are not enough for constructing the theory: they are enough for deriving finite bounds at all orders in renormalized perturbation theory, growing like $n!$ at the $n$-th order ({\it $n!$ bounds}), but the possible Borel summability of the series remains a  great challenge for the current century. A long term program addressing this issue has been started by Balaban-Feldman-Kn\"orrer-Trubowitz, see \cite{Balaban} for the state of the art of this project.  

Even though a full control of the corrections to Bogoliubov approximation is to date beyond reach of rigorous analysis, a few results are available if we focus on Bogoliubov predictions for the ground state energy. Indeed, mathematically, the validity of Bogoliubov's approach in three dimensional Bose gases has been first established by Lieb and Solovej for the computation of the ground state energy of bosonic jellium in \cite{LSo} and of the two-component charged Bose gas in \cite{LSo2} (upper bounds were  later given by Solovej in \cite{So}). Extending the ideas of \cite{LSo,LSo2}, Giuliani and Seiringer established in \cite{GiuS} the validity of the Lee-Huang-Yang formula (\ref{eq:LHY}) for Bose gases interacting through potentials scaling with the density to approach a simultaneous weak coupling and high density limit. This result has been recently improved by Brietzke and Solovej in \cite{BriS} to include a certain class of weak coupling and low density limits. It is worth to stress that in the regimes considered in \cite{LSo, LSo2, GiuS, BriS} the difference between first and second Born approximation and the full scattering length is small and it only gives negligible contributions to the energy. In other words, in the above mentioned regimes, cubic and quartic contributions neglected in Bogoliubov analysis can be proved to be small; this is crucial to make Bogoliubov's approach rigorous.  

An upper bound  for the ground state energy  in the thermodynamic limit coinciding with (\ref{eq:LHY}) up to second order was established in   \cite{YY} (improving a previous result by \cite{ESY}). Very recently, a lower bound for the ground state energy which establish the correct order of the next to leading order contribution in the whole dilute regime, has been obtained by \cite{BFS}, but without control on the constant.

Still, a proof of the validity of the predictions of Bogoliubov theory for the ground state energy in a regime of parameters where we cannot substitute the full scattering length with the first and second order terms in its Born approximation is missing. Moreover, no results concerning the excitation spectrum  are available in the thermodynamic limit.


\subsection{The Gross-Pitaevskii regime}

A natural question arising from the discussion of the previous section is whether some of the results predicted by Bogoliubov theory can be 
validated in regimes different from the thermodynamic limit, but still physically relevant for the description of Bose-Einstein condensates. This is the case of the so called {\it scaling regimes}, where the bosons are confined in a box of size length one and the interaction is allowed to depend on the number of particles. In the three dimensional case, it turns out interesting to  consider systems of $N$ bosons in the box 
$\Lambda = [-\frac12;\frac12]^{\times 3}$, described by the Hamilton operator
\begin{equation}\label{eq:ham-beta}  H_N^\beta = \sum_{j=1}^N -\Delta_{x_j} + \frac{\kappa}{N} \sum_{i<j}^N N^{3\beta} V (N^\beta (x_i -x_j)) \end{equation}
for a parameter $\beta \in [0;1]$, a coupling constant $\kappa > 0$ and a short range potential $V \geq 0$. Hamilton operators of the form (\ref{eq:ham-beta}) interpolate between the {\it mean-field regime} associated with $\beta = 0$ (effectively describing bosons interacting through weak and long range interactions) and the {\it Gross-Pitaevskii regime} corresponding to $\beta = 1$ (depicting a situation where interactions among the particles are strong and very short range). Note that, denoting with $\frak{a}_N$ the scattering length of the potential $N^{3\b-1} V(N^\b x)$,  for any $ \b \in [0,1]$ we have $\r \frak{a}^3_N = N^{-2}$, which corresponds to a diluteness condition.  Hence we may reasonably expect the predictions of Bogoliubov theory to hold for systems of bosons described by \eqref{eq:ham-beta}.





Since the Born series for the scattering length is an expansion in the ratio between the parameter  $\int V$ and the range of the potential, a simple computation shows that, in the regimes described by the Hamilton operator (\ref{eq:ham-beta}), replacing first and second Born approximations with the scattering length produces an error in the ground state energy of the order $N^{2\beta-1}$. Hence, one may guess that Bogoliubov truncation of the Hamiltonian can be rigorously justified for any $\b<1/2$. Indeed, starting from the pioneering work by \cite{Sei} several results have confirmed Bogoliubov picture in the mean-field limit $\beta =0$, both in the homogeneous and non homogeneous setting \cite{GS, LNSS, DN, LNR1, LNR2, P1, P2, P3}. On the other side for $\b\geq1/2$ Bogoliubov approximation fails. Nevertheless in \cite{BBCS2} the predictions of Bogoliubov theory where rigorously justified for any $0<\b<1$ (the proof in \cite{BBCS2} holds for $\k$ sufficiently small, but can be extended to any $\k$ using the strategy recently developed for the Gross-Pitaevskii regime in \cite{BBCS3}). The key idea to achieve this result was to understand the emergence of the scattering length as a consequence of correlations among the particles.

The Gross-Pitaevskii regime, where $\b=1$, is even more challenging from a mathematical point of view.  Indeed in this regime the ration between $\mathfrak{a}_N $ and the range of the potential is of order one, and all terms in the Born series of the scattering length contribute to the same order in $N$.
From a physical point of view the Gross-Pitaevskii regime owes its success to the fact that it represents a good description for the strong and short range interactions  among atoms in dilute, cold atomic gases. Moreover, it is the microscopic scaling leading to a rigorous derivation of the Gross-Pitaevskii equation for the dynamics of Bose-Einstein condensates (see \cite{BPS, BS}). Finally, it is easy to see that  $H_N^\b$ for $\b=1$ is equivalent to the Hamiltonian for  $N$ bosons in a box with $L=N$ interacting through a fixed potential $ V$; hence the Gross-Pitaevskii regime corresponds to a regime where the size of box is sent to infinity, but the system has in this limit a very low density $\rho= N/L^3 = N^{-2}$. 

It follows from the results of \cite{LSY,LS,LS2,NRS} that the ground state energy $E_N$ of the Gross-Pitaevskii Hamiltonian 
\begin{equation}\label{eq:HN}  
H_N = \sum_{j=1}^N -\Delta_{x_j} + \sum_{i<j}^N N^{2} V (N (x_i -x_j)) 
\end{equation}
defined on $L^2_s(\L^N)$ is such that 
\[
\lim_{N \to \infty} \frac{E_N}{N} = 4\pi \frak{a}_0 \, . \]
Furthermore, for any sequence of approximate ground states, ie. for any sequence $\psi_N \in L^2_s (\L^N)$ with $\| \psi_N \| = 1$ and  
\[
 \lim_{N \to \infty} \frac{1}{N} \langle \psi_N , H_N \psi_N \rangle = 4 \pi \frak{a}_0 \, , \]
the reduced density matrices $\gamma_N = \tr_{2, \dots , N} |\psi_N \rangle \langle \psi_N |$ are such that 
\begin{equation}\label{eq:BEC1} \lim_{N \to \infty} \tr \, \left| \gamma_N - |\ph_0 \rangle \langle \ph_0| \right| = 0 \end{equation}
where $\ph_0 \in L^2 (\Lambda)$ is the zero momentum mode defined by $\ph_0 (x) = 1$ for all $x \in \Lambda$. Aim of these notes it to discuss how to go beyond those result and computing the ground state energy and the low-lying excitation spectrum of (\ref{eq:HN}), up to errors vanishing in the limit $N \to \infty$. This is the content of  our main theorem.

\begin{theorem}\label{thm:main}
Let $V \in L^3 (\bR^3)$ be non-negative, spherically symmetric and compactly supported. Then, in the limit $N \to \infty$, the ground state energy $E_N$ of the Hamilton operator \eqref{eq:HN}
 is given by 
\begin{equation}
\begin{split}\label{1.groundstate}
E_{N} = \; &4\pi (N-1) \frak{a}_0 + e_\Lambda \frak{a}_0^2 \\ & - \frac{1}{2}\sum_{p\in\Lambda^*_+} \left[ p^2+8\pi \frak{a}_0  - \sqrt{|p|^4 + 16 \pi \frak{a}_0  p^2} - \frac{(8\pi \frak{a}_0 )^2}{2p^2}\right] + \cO (N^{-1/4}) \,,
    \end{split}
    \end{equation}     
with $\frak{a}_0$ the scattering length of $V$.  
 Here we introduced the notation $\Lambda_+^* = 2\pi \bZ^3 \backslash \{0 \}$ and we defined    
\[ 
   e_\Lambda = 2 - \lim_{M \to \infty} \sum_{\substack{p \in \bZ^3 \backslash \{ 0 \} : \\ |p_1|, |p_2|, |p_3| \leq M}} \frac{\cos (|p|)}{p^2} \]
where, in particular, the limit exists. Moreover, the spectrum of $H_N-E_{N}$ below a threshold $\zeta$ consists of eigenvalues given, in the limit $N \to \infty$, by 
    \begin{equation}
    \begin{split}\label{1.excitationSpectrum}
    \sum_{p\in\Lambda^*_+} n_p \sqrt{|p|^4+ 16 \pi \frak{a}_0  p^2}+ \cO (N^{-1/4} (1+ \zeta^3)) \, . 
    \end{split}
    \end{equation}
Here $n_p \in \bN$ for all $p\in\Lambda^*_+$ and $n_p \not = 0$ for finitely many $p\in \Lambda^*_+$ only.  
\end{theorem}

{\it Remarks:} 

\begin{itemize}
\item  By comparing \eqref{1.groundstate} and \eqref{1.excitationSpectrum}  with the predictions  \eqref{eq:BogEnergy} and \eqref{eq:BogSpec} from Bogoliubov theory, we see that in the Gross-Pitaevskii regime the integral of the potential is substituted by the scattering length $\frak{a}_0$, as physically expected. 
Indeed, if in (\ref{1.groundstate}) we replaced sums over discrete momenta $p \in \Lambda^*_+$ by integrals over continuous variables $p \in \bR^3$, we would obtain exactly (\ref{eq:LHY}). From this point of view Theorem \ref{thm:main} establishes the analog of the Lee-Huang-Yang formula for the ground state energy in the Gross-Pitaevskii regime. 

\item The term $e_\Lambda \frak{a}_0^2$ in (\ref{1.groundstate}) arises as a correction to 
the scattering length $\frak{a}_0$, due to the finiteness of the box $\Lambda$. For small interaction potentials, we can define a finite volume scattering length $\frak{a}_\L$ through the convergent Born series 
\[  8\pi \frak{a}_\L =   \widehat{V} (0)
+\sum_{k=1}^{\infty}\frac{(-1)^{k}}{(2N)^{k}} \sum_{p_1, \dots, p_{k}\in\Lambda^*_+ } \frac{\widehat{V} (p_1 /N)}{p_1^2} \left(\prod_{i=1}^{k-1}\frac{\widehat{V} ((p_i-p_{i+1})/N)}{p^2_{i+1}}\right)  \widehat{V} (p_{k}/N) \, .  \]
In this case, one can check that   
\[ \lim_{N \to \infty} 4\pi (N-1) \left[ \frak{a}_0 -  \frak{a}_\L \right] = e_\Lambda \frak{a}_0^2 \, .  \] 
Observe that, if we replace the potential $V$ by a rescaled interaction $V_R (x) = R^{-2} V(x/R)$ with scattering length $\frak{a}_R = \frak{a}_0 R$ then, for large $R$ (increasing $R$ makes the effective density larger), 
the order one contributions to the ground state energy scale as $e_\Lambda \frak{a}_0^2 R^2$ and, respectively, as 
\begin{equation}\label{eq:Bog-sum} \begin{split} -\frac{1}{2} \sum_{p \in 2\pi \bZ^3 \backslash \{ 0 \}}  &\left[ p^2 + 8\pi \frak{a}_0 R - \sqrt{|p|^4 + 16 \pi \frak{a}_0 R p^2} -  \frac{(8 \pi \frak{a}_0 R)^2}{p^2} \right] \\ = \; &\frac{R}{2} \sum_{p \in \frac{2\pi}{\sqrt{R}}  \bZ^3 \backslash \{ 0 \}} \left[ p^2 + 8\pi \frak{a}_0 - \sqrt{|p|^4 + 16 \pi \frak{a}_0 p^2} -  \frac{(8 \pi \frak{a}_0)^2}{p^2} \right]  \\
\simeq \; &\frac{R^{5/2}}{2(2\pi)^3} \int_{\bR^3} \left[ p^2 + 8\pi \frak{a}_0 - \sqrt{|p|^4 + 16 \pi \frak{a}_0 p^2} -  \frac{(8 \pi \frak{a}_0)^2}{p^2} \right]  dp \\ = \; & \frac{4\pi R^{5/2} (16 \pi \frak{a}_0)^{5/2}}{15 (2\pi)^3} \, = 4\pi \frak{a}_0 \cdot \frac{128}{15 \sqrt{\pi}} \frak{a}_0^{3/2} R^{5/2} . \end{split} \end{equation}
In particular, letting $R \to \infty$ (independently of $N$), it follows the finite volume correction becomes subleading, compared with (\ref{eq:Bog-sum}). 
\item Theorem \ref{thm:main} gives precise information on the low-lying eigenvalues of (\ref{eq:HN}). The approach in \cite{BBCS4} combined with standard arguments, also gives information on the corresponding eigenvectors. In \cite{BBCS4} we provide a norm approximation of eigenvectors associated with the low-energy spectrum of (\ref{eq:HN}). As an application, we can compute the  condensate depletion in the ground state $\psi_N$ of (\ref{eq:Ham0}), confirming Bogoliubov prediction.

\end{itemize}

In the rest of these notes, we are going to describe the strategy leading to Theorem \ref{thm:main}, as obtained in \cite{BBCS3} and \cite{BBCS4}. In particular we will focus on the new ideas which are needed to remove the assumption of small interaction potential that was previously used in \cite{BBCS1}. 

\section{A Fock space representation for excited particles}   \label{Sec:LN}

The first step in the proof of Theorem \ref{thm:main} consists in a rigorous version of the substitution  of the creation and annihilation operators in the condensate by scalar numbers, which is the first step in Bogoliubov theory. Following an idea from \cite{LNSS}, we use the Fock space to describe orthogonal excitations with respect to the condensate wave function\footnote{Usually the second quantization formalism is used to represent the system in a grand-canonical picture, where the particle number can vary. On the contrary here the particle number is fixed to be $N$, but the excitation number is not fixed, and can vary up to $N$.}. 
More precisely, we write any arbitrary $N$-particle wave function $\ps \in L^2_s(\L^N)$ as
\be \label{eq:trialstate1}
\psi_N = \alpha_0 \ph_0^{\otimes N} + \alpha_1 \otimes_s \ph_0^{\otimes (N-1)} + \dots + \alpha_N 
\ee 
with $\alpha_j \in L^2_{s,\perp} (\Lambda^j)$ for all $j=0,1, \ldots, N$. Here $\ph_0(x)$ is the condensate wave function in \eqref{eq:BEC1}. Moreover $L^2_\perp (\Lambda)$ denotes the orthogonal complement of the one-dimensional subspace spanned by $\ph_0 $ in $L^2(\L)$, and $ L^2_{s,\perp} (\Lambda^j)$ the symmetric tensor product of $j$ copies of  $L^2_\perp (\Lambda)$.
It is easy to check that the decomposition  \eqref{eq:trialstate1} defines a unitary map $U_N$ from the space $ \in L^2_s(\L^N)$ to the truncated Fock space constructed over $L^2_\perp (\Lambda)$
\[
\cF^{\leq N}_+ = \bigotimes_{j=0}^N L^2_{s,\perp} (\Lambda^j)\,.
\]
For a $\psi_N \in L^2_s(\L^N)$ we denote with $\xi_N$ the corresponding excitation vector
\[
\xi_N := U_N \psi_N = \{\a_0, \a_1, \ldots, \a_N \} \in \cF^{\leq N}_+ \,.
\]
The action of the unitary operator $U_N$ on products of a creation and an annihilation operator (products of the form $a_p^* a_q$ can be thought of as operators mapping $L^2_s (\Lambda^N)$ to itself) is reminiscent of Bogoliubov substitution. Indeed, for any $p,q \in \Lambda^*_+ = 2\pi \bZ^3 \backslash \{ 0 \}$, we find (see \cite{LNSS}):
\begin{equation}\label{eq:U-rules}
\begin{split} 
U_N \, a^*_0 a_0 \, U_N^* &= N- \cN_+  \\
U_N \, a^*_p a_0 \, U_N^* &= a^*_p \sqrt{N-\cN_+ } \\
U_N \, a^*_0 a_p \, U_N^* &= \sqrt{N-\cN_+ } \, a_p \\
U_N \, a^*_p a_q \, U_N^* &= a^*_p a_q \,,
\end{split} \end{equation}
with $\cN_+= \sum_{p \in \Lambda^*_+} a^*_p a_p$ the operator counting the number of excited particles (here $a^*_p$ and $a_p$ are the usual creation and annihilation operators defined on the bosonic Fock space $\cF = \otimes_{j\geq0} L^2(\Lambda)^{\otimes_s j}$ and satisfying canonical commutation relations $[a_p, a^*_q] = \d_{pq}$ and $[a_p, a_q]= [a^*_p, a^*_q]=0$)\,.

Using $U_N$, we can define an excitation Hamiltonian $\cL_N := U_N H_N U_N^*$, acting on a dense subspace of $\cF_+^{\leq N}$. To compute the operator $\cL_N$, we first write the Hamiltonian (\ref{eq:HN}) in momentum space, in terms of creation and annihilation operators. We find 
\begin{equation}\label{eq:Hmom} H_N = \sum_{p \in \Lambda^*} p^2 a_p^* a_p + \frac{1}{2N} \sum_{p,q,r \in \Lambda^*} \widehat{V} (r/N) a_{p+r}^* a_q^* a_{p} a_{q+r} 
\end{equation}
where 
\[ \widehat{V} (k) = \int_{\bR^3} V (x) e^{-i k \cdot x} dx \] 
is the Fourier transform of $V$, defined for all $k \in \bR^3$. Using \eqref{eq:U-rules} we conclude that
\begin{equation}\label{eq:cLN} \cL_N =  \cL^{(0)}_{N} + \cL^{(2)}_{N} + \cL^{(3)}_{N} + \cL^{(4)}_{N} \end{equation}
with
\begin{equation*}\label{eq:cLNj} \begin{split} 
\cL_{N}^{(0)} =\;& \frac{N-1}{2N}\widehat{V} (0) (N-\cN_+ ) + \frac{\widehat{V} (0)}{2N} \cN_+  (N-\cN_+ ) \\
\cL^{(2)}_{N} =\; &\sum_{p \in \Lambda^*_+} p^2 a_p^* a_p + \sum_{p \in \Lambda_+^*} \widehat{V} (p/N)  a_p^* a_p \left( \frac{N-\cN_+}{N} \right)  \\ 
&+ \frac{1}{2} \sum_{p \in \Lambda^*_+} \widehat{V} (p/N) \left[ a_p^* a_{-p}^*  \sqrt{\frac{N-1-\cN_+}{N} \frac{N-\cN_+}{N}} + \hc \right] \\
\cL^{(3)}_{N} =\; &\frac{1}{\sqrt{N}} \sum_{p,q \in \Lambda_+^* : p+q \not = 0} \widehat{V} (p/N) \left[ a^*_{p+q} a^*_{-p} a_q \sqrt{\frac{N-\cN_+}{N}} + \hc \right] \\
\cL^{(4)}_{N} =\; & \frac{1}{2N} \sum_{\substack{p,q \in \Lambda_+^*, r \in \Lambda^*: \\ r \not = -p,-q}} \widehat{V} (r/N) a^*_{p+r} a^*_q a_p a_{q+r} \,.
\end{split} \end{equation*}
As in Bogoliubov theory, conjugation with $U_N$ extracts, from the original quartic interaction, some constant, quadratic and cubic contributions, collected in $\cL^{(0)}_N$, $\cL^{(2)}_N$ and $\cL^{(3)}_N$ respectively. The challenge of the Gross-Pitaevskii regime is that, due to the slow decay of $\widehat V(p/N)$ for large momenta, we cannot neglect the cubic and quartic contributions in \eqref{eq:cLN} for $N \to \io$. This fact can be understood from different point of views.

\begin{itemize}
\item It is well known that the ground state energy of bosons in the Gross-Pitaevskii regime are characterized by a correlation structure which varies on the length scale of the scattering length of the interaction $\frak{a}_N \sim N^{-1}$, and which can be modeled by solution of the zero energy scattering equation. This is the key ingredient to show upper and lower bounds consistent with \eqref{1.groundstate} at leading order \cite{LSY},  and to establish the results in \cite{BFS, ESY,YY}. The same correlation structure has to be included in any approach aimed to show the emergence of the Gross-Pitaevskii equation as an effective description for the evolution of initially trapped Bose-Einstein condensates which evolves under the dynamics generated by \eqref{eq:HN} (see \cite{BPS} and references therein). On the contrary, the application of the unitary map $U_N$ only factors out the condensate, but does not remove the short scale correlation structure that, as we will see below, still carries an energy of order $N$.

%


\item From a renormalization group perspective studying $\cL_N$ corresponds to carry out perturbation theory around Bogoliubov approximation for momenta larger than $2\pi$. However, as already commented in the introduction, such a theory is divergent in the ultraviolet, and to get a well defined theory we need to renormalize both the quadratic and the cubic vertices of the theory. 

\end{itemize}


Before describing how to include the correlation structure into our analysis, let us explain the guiding idea behind our overall strategy, which can be easily illustrated  in the simpler case were we substitute $\widehat V(p/N)$ by a mean-field potential $\kappa \widehat V(p)$ with intensity $\kappa>0$ sufficiently small. 

\subsection{A sketch of the strategy in the mean field case} \label{sec:MF}
 In the approach we are going to follow, the key ingredient used to investigate the validity of Bogoliubov theory is the proof of optimal bounds on the number and energy of excitations in low energy states.  With this goal in mind, let us denote with $\cL_N^{\text{mf}}$ an excitation Hamiltonian identical to \eqref{eq:cLN} except that $\widehat V(p/N)$ is substituted by $\kappa \widehat V(p)$. It is easy to check that there exists a constant  $C >0$ such that
\be \label{eq:lowMF} \begin{split}
\cL_N^{\text{mf}} &\geq \frac{N}{2} \widehat V(0)  + \sum_{p \in \L_+^*} p^2 a^*_p a_p  \\
& \qquad +  \frac{\k}{2N} \sum_{\substack{p,q \in \Lambda_+^*, r \in \Lambda^*: \\ r \not = -p,-q}} \widehat{V} (r) a^*_{p+r} a^*_q a_p a_{q+r} -  C  \kappa \, (\cN_+ +1)\,,
\end{split}\ee
where we used that $\cN_+ \leq N$. Using  positivity of the interaction and the gap in the kinetic energy $ \sum_{p \in \L_+^*} p^2 a^*_p a_p  \geq (2 \pi)^2 \cN_+$, we obtain that for sufficiently small $\k$ there exists $C>0$ such that
\be  \label{eq:lowerMF}
\cL_N^{\text{mf}} \geq   \frac{N}{2} \widehat{V}(0) + c \, \cN_+- C\,.
\ee
This also implies the lower bound  $\cL_N^{\text{mf}} \geq \frac{N}{2} \widehat V (0) + C$. 
On the other side by using the vacuum state in $\cF^{\leq N}_+$ as a trial state we obtain the upper bound 
\[
\cL_N^{\text{mf}} \leq  \frac{N}{2} \widehat{V}(0) \,.
\]
This allows us to conclude  that the ground state energy of our mean-field Hamiltonian
\[
 H_N^{\text{mf}} = \sum_{p \in \Lambda^*} p^2 a_p^* a_p + \frac{\k}{2N} \sum_{p,q,r \in \Lambda^*} \widehat{V} (r) a_{p+r}^* a_q^* a_{p} a_{q+r} 
\]
 satisfies the bound $|E_N^{\text{mf}} -  N\widehat V(0)/2 |\leq C$. Moreover for any $N$-particle wave function $\ps_N \in L^2_s(\L^N)$ such that 
\be \label{eq:lowenstate}
\bmedia{\ps_N, H_N^{\text{mf}} \ps_N} \leq \frac{N}{2} \widehat{V}(0)  + \zeta
\ee
we have that the corresponding excitation vector $\xi_N = U^* \ps_N$ satisfies
\[
\bmedia{\xi_N, \cL_N^{\text{mf}} \xi_N} \leq \frac{N}{2} \widehat{V}(0)  + \zeta\,,
\]
and hence, through \eqref{eq:lowerMF},
\[
\bmedia{\xi_N, \cN_+ \xi_N} \leq C ( 1+\zeta) \,.
\]
Hence low energy states have a bounded number of excitations. 

Additionally, denoting 
\[
\cK= \sum_{p \in \L_+^*} p^2 a^*_p a_p\]
the kinetic energy of excitations, and with
\[
\cV_N^{\text{mf}}= \frac{\k}{2N} \sum_{\substack{p,q \in \Lambda_+^*, r \in \Lambda^*: \\ r \not = -p,-q}} \widehat{V} (r) a^*_{p+r} a^*_q a_p a_{q+r}\]
the potential energy in the mean-field scaling, from \eqref{eq:lowMF} we also obtain 
\be \label{eq:lowMF2}
\cL_N^{\text{mf}} \geq   \frac{N}{2} \widehat{V}(0)  + \frac 12 (\cV_N^{\text{mf}} + \cK ) - C\,.
\ee 
This implies that any excitation vector $\xi$ associated to low energy states in the sense of \eqref{eq:lowenstate} has a bounded excitation energy, namely satisfies
\[
\bmedia{\xi_N, \cH_N^{\text{mf}} \xi_N} \leq C ( 1+\zeta)  \,,
\]
with $\cH_N^{\text{mf}} :=\cK +\cV_N^{\text{mf}} $.  

We can derive even stronger bounds on the excitation vector $\xi_N$ associated with a normalized $N$-particle wave function $\psi_N$, if instead of imposing the condition  \eqref{eq:lowenstate}, we require $\psi_N$ to belong to the spectral subspace of $H_N$ associated with energies below  $\frac{N}{2} \widehat{V}(0) +\zeta$. To this aim we define 
\[\tl \cL_N = \cL_N^{\text{mf}} -\frac{N}{2} \widehat{V}(0)\,. \]
Then
\[ \begin{split}
\bmedia{\xi_N, \cN_+ \cH_N^{\text{mf}} \xi_N} =\; &  \bmedia{\xi_N, \cN_+^{1/2} \cH_N^{\text{mf}}  \cN_+^{1/2}\xi_N}  \\
\leq \; &  \bmedia{\xi_N, \cN_+^{1/2} \tl \cL_N\, \cN_+^{1/2}\xi_N} \\
\leq \; &  \bmedia{\xi_N, \cN_+^{1/2} [\tl \cL_N,  \cN_+^{1/2}]\, \xi_N} +  \bmedia{\xi_N, \cN_+ \tl \cL_N \xi_N} \\
\end{split}\]
where in the second line we used \eqref{eq:lowMF2}. Using the assumption of $\xi_N$ being in the spectral subspace of $\tl \cL_N$ associated with energies below $\zeta$ we get
\[\begin{split}
\bmedia{\xi_N, \cN_+ \tl \cL_N \xi_N} &\leq  \bmedia{\xi_N, \cN_+  \xi_N}^{1/2} \bmedia{\tl \cL_N \xi_N, \cN_+\tl \cL_N\xi_N}^{1/2} \\
&\leq  C (1+\zeta) \, \bmedia{ \tl \cL_N \xi_N, \cH_N^{\text{mf}} \tl \cL_N\xi_N}^{1/2}  \\
& \leq  C (1+\zeta^2) \,.
\end{split}\]
On the other side, to bound the term $  \bmedia{\xi_N, \cN_+^{1/2} [\tl \cL_N,  \cN_+^{1/2}]\, \xi_N}$ we uses that the commutator $[\tl \cL_N,  \cN_+^{1/2}]$ can be computed explicitly. More precisely one can show that the operator $A=(\cH_N^{\text{mf}}+1)^{-1/2} [\tl \cL_N,  \cN_+^{1/2}] (\cH_N^{\text{mf}}+1)^{-1/2}$ is  a self-adjoint operator on $\cF^{\leq N}_+$ whose norm is bounded uniformly in $N$, this leading to the bound
\[ \begin{split}
 \bmedia{\xi_N, \cN_+^{1/2} [\tl \cL_N,  \cN_+^{1/2}]\, \xi_N} &\leq  \bmedia{\xi_N, \cN_+ (\cH_N^{\text{mf}}+1) \xi_N}^{1/2}   \bmedia{\xi_N,  (\cH_N^{\text{mf}}+1) \xi_N}^{1/2} \\
& \leq \d  \bmedia{\xi_N,  \cN_+ \cH_N^{\text{mf}}\xi_N} + C (1+ \zeta)\,.
\end{split}
\]
We conclude that 
\[
\bmedia{\xi_N, \cN_+ \cH_N^{\text{mf}} \xi_N} \leq C (1+\zeta^2)\,.
\]
By induction similar bounds can be proved for expectations of products of the form  $( \cH_N^{\text{mf}}+1)(\cN_++1)^k $ onto excitation vectors which are in the spectral subspace of $\tl \cL_N$ associated with energy below $\zeta$, for any $k \in \bN $. 

Armed with this stronger bounds one can analyze the excitation Hamiltonian $\cL_N^{\text{mf}}$ from a different perspective,  and show that the cubic and quartic terms in $\cL_N^{\text{mf}}$ are bounded by $C N^{-1/2} (\cN+1)^2$, and are therefore negligible on low energy states, according to Bogoliubov picture.  \\


If we now want to apply the strategy sketched above to the Gross-Pitaevskii regime, already in the simpler case of sufficiently small unscaled potential, we find two main difficulties.  First of all the ground state energy in the Gross-Pitaevskii regime is given at  leading order in $N$ by $4 \pi \frak{a}_0 N$, which is strictly smaller than $ \frac{N}{2} \widehat{V}(0) $. Moreover the quadratic non diagonal term are large (of order $N$), due to the slow decay of the interaction. Both problems are related to the fact that we need to extract from $\cL_N^{(3)}$ and $\cL_N^{(4)}$ important contributions to the energy of low energy states. This is what we are going to describe in the next section.


\section{Correlations between condensate and excitation pairs}    \label{sec:GN}

In the last section we emphasized many times that in order to deal with the Gross-Pitaevskii regime we should take into account correlations among particles.
We include correlations in $\cF^{\leq N}_+$ by means of a suitable unitary operator which models the creation (annihilation) of excitation pairs out of the condensate. The idea of factoring out correlations using unitary operators in the Fock space (and in particular Bogoliubov transformations) dates back to \cite{BDS}.  In our setting, to make sure that the truncated Fock space $\cF^{\leq N}_+$ remains invariant, we will have to use generalized Bogoliubov transformation. For $\mu>0$ we define the operator
\be \label{eq:defT}
T(\eta_H) = \exp \left[ \frac 12 \sum_{|p|\geq \mu}  \eta_p \big( b^*_p b^*_{-p} - b_p b_{-p})\right]\,,
\ee
where we introduced generalized creation and annihilation operators 
\[ 
b^*_p = a^*_p \, \sqrt{\frac{N-\cN_+}{N}} , \qquad \text{and } \quad  b_p = \sqrt{\frac{N-\cN_+}{N}} \, a_p 
\]
for all $p \in \Lambda^*_+$. To understand the role of the $b_p$ and $b^*_p$ operators is sufficient to observe that, by (\ref{eq:U-rules}), 
\[ U_N^* b_p^* U_N = a^*_p  \frac{a_0}{\sqrt{N}}\,, \qquad U_N^* b_p U_N = \frac{a_0^*}{\sqrt{N}} a_p \,.\]
In other words $b_p^*$ creates a particle with momentum $p \in \Lambda^*_+$ but, at the same time, it annihilates a particle from the condensate; it creates an excitation, preserving the total number of particles in the system. This guarantees that $T(\eta_H)$ is an operator from $\cF^{\leq N}_+$ into itself. The  action of $T(\eta_H)$ on $b_p$ and $b^*_p$ is reminiscent of the action of a usual Bogoliubov transformation (which would be defined as \eqref{eq:defT}, but with usual creation and annihilation operators). Indeed, for any p such that $|p| \geq \mu$ we have
\be \begin{split} \label{eq:actT}
T^*(\eta_H) b_p T(\eta_H) &= \cosh(\eta_p) b_p + \sinh(\eta_p) b^*_{-p} + d_p \\
T^*(\eta_H) b_p^* T(\eta_H) &= \cosh(\eta_p) b^*_p + \sinh(\eta_p) b_{-p} + d^*_p\,,
\end{split}\ee
where it is possible to prove that  the operators $d_p$ and $d_{-p}$ produce small contributions on states with a bounded number of excitations, as those we are going to consider. We refer the reader to \cite[Lemma 3.4]{BBCS3} for the precise estimates satisfied by the $d_p$ operators and a discussion of this point. For the sake of these notes it will be sufficient to think to the operators $b_p$ and $b^*_p$ as acting as usual creation/annihilation operators. 


It remains to discuss the role of the function $\eta_p$  appearing in \eqref{eq:defT}. The correct choice for this function results to be
\be \label{eq:defeta}
\eta_p = \frac 1 {N^2} \widehat{(1-f_N)}(p/N)\,,
\ee
with $f_N(x)$ the solution of the Neumann problem
\[
\Big( - \D + \frac 12 N^2 V(Nx) \Big) f_N(x) = \l_N f_N(x)
\]
on the ball $|x|\leq 1/2$, with $f_N(x)=1$ and $\dpr_{|x|} f_N(x)=0$ for $|x|=1/2$. The function $f_N(x)$ is a slight modification of the zero-energy infinite volume scattering length and in particular we have
\be \label{eq:intVf}
\Big| \int N^3 V(Nx) f_N(x) - 8 \pi \frak{a}_0 \Big| \leq \frac  {C \frak{a}_0^2} N \,,
\ee
to be compared with \eqref{eq:fraka0}. The properties of  $f_N(x)$ can be found in  \cite[Lemma 3.1]{BBCS3}. What is relevant for the next analysis is that as a consequence of the definition \eqref{eq:defeta} we have
\[ 
|\eta_p|\leq \frac{k}{|p|^2} e^{-|p|/N}\,.
\]
Hence, defining \[ 
\label{eq:defetaH}
\eta_H(p)= \eta_p \chi(|p| \geq \mu)\] we have $\| \eta_H\|_2 \leq C \mu^{-1}$ (in the following $\mu^{-1}$ will play the role of the small parameter in the general situation where the unscaled potential is not small).  Note that conjugation by $T(\eta_H)$ does not change substantially the number of excitations. Indeed by using \eqref{eq:actT} it is easy to check that the number of excitations on a excitation state $\xi_N = T(\eta) \O_N$, with $\O_N=\{1, 0,0, \ldots, 0\}\in \cF^{\leq N}_+$ the vacuum state, is given by
\[
\bmedia{T(\eta_H) \O_N, \cN_+ T(\eta_H) \O_N} \leq  C \| \eta_H \|  \,.
\]
More in general one can show the following lemma (see  \cite[Lemma 3.1]{BS}):
\begin{lemma}\label{lm:Ngrow}
For every $n \in \bN$ there exists a constant $C > 0$ such that, on $\cF_+^{\leq N}$, 
\[\label{eq:bd-Beta} T^*(\eta_H)(\cN_+ +1)^{n} T(\eta_H) \leq C e^{C \| \eta_H \|} (\cN_+ +1)^{n}  
\]
\end{lemma}
 On the other side, since $\| \eta\|_{H^1}\leq C \sqrt N$, we expect the few excitations that we are introducing through $T(\eta_H)$ to carry a large (order $N$) contribution to the energy. Therefore conjugation with $T(\eta_H)$ has a chance 
to decrease the vacuum expectation of the excitation Hamiltonian $\cL_N$ to $4\pi \frak{a}_0 N$ (to leading order).  With this motivation in mind, we define a new excitation Hamiltonian $\cG_{N} : \cF^{\leq N}_+ \to \cF^{\leq N}_+$ by setting 
\[ 
\cG_{N} = T^*(\eta_H) \cL_N T(\eta_H) = T^*(\eta_H) U_N H_N U_N^* T(\eta_H) \,.
\]
The outcome of the action of $T(\eta_H)$ on $\cL_N$ is summarized by the next proposition, which was proved in \cite{BBCS4}. 

\begin{prop} \label{prop:GN}
Let $V \in L^3(\bR^3)$ be compactly supported, pointwise non-negative and
spherically symmetric. Then
\begin{equation}\begin{split} \label{eq:GNeff} \cG_{N}  =\; & 4\pi \frak{a}_0 N +\cH_N  \\[3pt]
&  + [2\widehat V(0)- 8 \pi \frak{a}_0] \sum_{p \in \L^*_+ :\, |p| \leq \mu}   a^*_pa_p (1-\cN_+/(2N) )  \\
		&   + 4\pi \frak{a}_0\sum_{p \in \L^*_+ :\, |p| \leq \mu}   \big[ b^*_p b^*_{-p} + b_p b_{-p} \big] \\
		& +  \frac{1}{\sqrt N}\sum_{p,q\in\Lambda_+^*: p+q\neq 0} \widehat V(p/N)\big[ b^*_{p+q}a^*_{-p}a_q+ \emph{h.c.}\big] + \cE_{\cG_{N}} \,,
		\end{split}\end{equation} 
where there exist constants $C, \a, \b > 0$ such that 
\[\label{eq:GeffE} \pm \cE_{\cG_{N}} \leq  \frac{C}{\mu^{\a}} \cH_N + C \mu^{\b}\,. \]
\end{prop}

We see that indeed the action of $T(\eta_H)$ renormalizes the constant part of the energy (at leading order) and the non diagonal quadratic contributions. Let us quickly explain the mechanism behind the outcome of Prop.\ref{prop:GN}. Writing $T = e^{B(\eta_H)}$, with $B(\eta_H) = (1/2) \sum_{p \in \Lambda^*_+} \eta_H(p)  \left( b_p^* b_{-p}^* -  b_p b_{-p} \right)$, we observe that 
\be \label{eq:GNexp} \begin{split}
\cG_{N} = \, & T^*(\eta_H) \cL_N T(\eta_H)  = e^{-B(\eta_H) } \cL_N e^{B(\eta_H) } \\
& \simeq \cL_N + [ \cL_N, B(\eta_H)  ] + \frac{1}{2} [[ \cL_N , B(\eta_H)  ] , B(\eta_H)  ]  + 
\dots \, .  \end{split}\ee
The commutator $[\cL_N, B(\eta_H) ]$ contains the contributions $[\cK , B(\eta_H) ]$ and $[\cV_N , B(\eta_H) ]$. Up to small errors, we find
\begin{equation}\label{eq:cK-com} [ \cK, B(\eta_H)  ] \simeq \sum_{\substack{p \in \L^*_+\\ |p| \geq \mu}} p^2 \eta_p \left[ b_p^* b_{-p}^* + b_p b_{-p} \right] \end{equation}
and
\begin{equation}\label{eq:cVN-com} [\cV_N, B(\eta_H) ] \simeq \frac{1}{2N} \sum_{\substack{p,q \in \L^*_+\\ |p+q| \geq \mu}} \widehat{V} (q/N) \eta_{q+p} \left[ b_p^* b_{-p}^* + b_p b_{-p} \right]  \, .  \end{equation}
In fact, the commutator $[\cV_N, B(\eta_H) ]$ is approximately quartic in creation and annihilation operators. Rearranging it in normal order, however, we obtain the quadratic contribution (\ref{eq:cVN-com}) (the remaining, normally ordered, quartic term is negligible). With the appropriate choice of the coefficients $\eta_p$ (given by \eqref{eq:defeta}), we can combine the large term
\[
\frac 12 \sum_{p \in \L^*_+} \widehat{V}(p/N) \left[ b^*_p b^*_{-p} + \hc \right]
\]
with (\ref{eq:cK-com}), (\ref{eq:cVN-com}), so that their sum can be estimated by $C \| \eta_H \| (\cN_++1)$. At the same time, the second commutator $[[ \cL_N , B(\eta_H)  ] , B(\eta_H)  ]$ produces new constant terms that, again with the choice \eqref{eq:defeta} of $\eta_p$, change the vacuum expectation to its correct value $4\pi \frak{a}_0 N$. 

An important remark is that conjugation by $T(\eta_H)$ leaves the cubic term and the quadratic diagonal terms in $\cL_N$ unchanged. For interactions $\k V$  with sufficiently small intensity $\k>0$ we can bound all the diagonal quadratic terms by $C \k (\cH_N+1)$. 
Moreover, after writing the cubic term in position space we get
\[ \begin{split}
& \big| \bmedia{\xi, \frac{1}{\sqrt N} \int dx dy N^3 \k V(N(x-y)) b^*_x a^*_y a_x \xi} \big | \\
& \quad  \leq \Big[  \int dx dy N^2 \k V(N(x-y)) \| a_x a_y \xi \|^2 \Big]^{1/2}  \Big[  \int dx dy N^3 \k V(N(x-y)) \| a_x  \xi \|^2 \Big]^{1/2}  \\
& \quad \leq C \k^{1/2} \bmedia{\xi, \cV_N \xi}^{1/2} \bmedia{\xi, \cN_+\xi}^{1/2}\\
& \quad  \leq C \k^{1/2} \bmedia{\xi, \cH_N \xi}\,.
\end{split}
\]
Hence, for weak interaction potentials Prop.\ref{prop:GN} immediately implies the lower bound 
\[
\cG_{N} \geq  4 \pi \frak{a}_0 N + \frac 12 \cH_N  - C_{\mu, \k} \,\cN_+ -C \,,
\]
where the constant $C_{\mu, \k}>0$ can be chosen to be sufficiently small by choosing $\mu^{-1}$ and $\k$ sufficiently small. Hence the error term proportional to the number of particles operator can be controlled by the gap in the kinetic energy, and one can repeat for the Gross-Pitaevskii interaction the same strategy sketched in Sec.\ref{sec:MF}. This is the approach used in \cite{BBCS1} to show  condensation with optimal rate for bosons in the Gross-Pitaevskii regime,  under the assumption of small unscaled potential. \\

For large potentials it is clear that conjugation by $T(\eta_H)$ is not enough to take advantage of the kinetic energy gap. In fact we can only show the following proposition (see \cite{BBCS3} for a proof).
\begin{prop} Let $V \in L^3(\bR^3)$ be compactly supported, pointwise non-negative and
spherically symmetric. Then
 \[ \label{eq:GNell-prel}
\cG_{N}  = 4 \pi \frak{a}_0 N + \cH_N + \theta_{\cG_{N}}
\]
where for every $\delta > 0$ there exists constants $C, \a > 0$ such that  
\[
\pm \theta_{\cG_{N}} \leq \delta \cH_N + C \mu^{\a} (\cN_+ + 1)
\]
and the improved lower bound 
\be  \label{eq:Gbd0}
\theta_{\cG_{N}} \geq  - \delta \cH_N - C \cN_+ - C \mu^\beta 
\ee
hold true for $\mu$ (of order one) sufficiently large and $N \in \bN$ large enough.  
\end{prop}

The remaining part of these notes are devoted to explain how to extend our analysis to large potentials. Looking at \eqref{eq:GNeff} it might appear evident that one possible route to this extension is to take into account for additional correlations, to renormalize the cubic term on the r.h.s. of \eqref{eq:GNeff}. Mathematically, this is achieved by conjugating $\cG_{N}$  with an additional unitary operator, given by the exponential of an operator cubic in creation and annihilation operators, as described in the following section. 


\section{Correlations due to triplets}

\begin{figure}[t] 
\centering
\begin{tikzpicture}[scale =0.6,
    every path/.style = {},
 ]
  \begin{scope}
\draw[thick,|-] (0,1.5) -- (4,1.5);
\draw[thick, |-] (4,1.5) -- (11, 1.5);
\draw[thick, ->] (11,1.5) -- (12, 1.5);  
\node[below] at (0,0.2) {$0$};
\node[below] at (4.5,0.4) {$ (16 \pi \mathfrak{a}_0)^{1/2}$};
\node[above] at (13,1) {$|p|$};
\draw[ |->, thick] (0,3.2) -- (4, 3.2);
\draw[ <-, thick] (4,3.2) -- (12, 3.2);
\node[above] at (8,3.5) {\scriptsize free particle regime};
\node[above] at (2,3.5) {\scriptsize linear spectrum};
\draw[-, color=orange, very thick, dashed] (0.4,2.4) -- (5.5,2.4);
\draw[-, color=orange, very thick, dashed ] (0.4,0.6) -- (5.5,0.6);
\draw[-, color=orange, very thick, dashed] (5.5,0.6) -- (5.5,2.4);
\draw[-, color= orange, very thick, dashed] (0.4,0.6) -- (0.4,2.4);
\node[below] at (2, 1.5) {\textcolor{orange}{$P_L$}};
\node[above] at (5.8,2.3) {\textcolor{orange}{$\nu$}};
1
\draw[-, color=violet, very thick, dashed] (7,2.4) -- (12,2.4);
\draw[-, color=violet,  very thick, dashed ] (7,0.6) -- (12,0.6);
\draw[-, color=violet,  very thick, dashed] (7,0.6) -- (7,2.4);
\node[below] at (9, 1.5) {\textcolor{violet}{$P_H$}};
\node[above] at (6.8,2.2) {\textcolor{violet}{$\mu$}};
\end{scope}
\end{tikzpicture}
\centering \begin{minipage}{12cm}
\caption{\small Schematical picture of the high and low momenta sets entering in the definition of the cubic operator $A$ defined in \eqref{eq:Aell1}. The arrowed line represents the energy scale of the problem going from zero energy to high energy (ultraviolet).  There are two energy scales in our problem: the first is the inverse of the range of the potential, which in our case is of order $N$; the second is provided by the expected value of the velocity of sound, equal to $\sqrt{16 \pi \frak{a}_0}$, which is in our setting of order one. The latter corresponds to the scale below which the low energy excitation spectrum behaves linearly.    } \label{Fig1}

\end{minipage}
\end{figure}
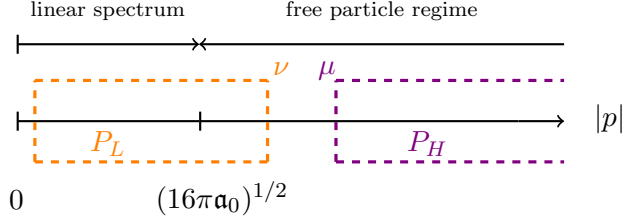

To renormalize the cubic term on the r.h.s. of \eqref{eq:GNeff} we include correlations due to triplets. For a parameter $0< \nu < \mu$ we define the low-momentum set 
\[  P_{L} = \{p\in \Lambda_+^*: |p| \leq \nu \}\,. \]
Notice that the high-momentum set entering in the quadratic operator $T(\eta_H)$
\[ 
P_H = \{p\in \Lambda_+^*: |p| \geq \mu \} 
\]
and $P_{L}$ are separated by a set of intermediate momenta $\nu < |p| < \mu$. We introduce the operator $A : \cF_+^{\leq N} \to \cF_+^{\leq N}$, by 
\be \label{eq:Aell1}
A = \frac{1}{\sqrt N} \sum_{\substack{ r \in P_H \\ v \in P_L }} \eta_r \big [\; b^*_{r+v} a^*_{-r} a_v  - \hc \; \big]\,.
\ee
While the generalized Bogoliubov transformation $T(\eta_H)$ used in the definition of $\cG_{N}$ described scattering processes involving two excitations with momenta $p$ and $-p$ and two particles in the condensate (i.e. two particles with zero momentum),  the cubic operator $A$ corresponds to processes involving two excitations with large momenta $p$ and $p+v$, an excitation with small momentum $v$, and a particle in the condensate.   Here large and small refer to the expected value of the sound velocity $\sqrt{16 \pi\frak{a}_0}$, which represents the separation between momenta for which we expect a linear spectrum of excitations and momenta for which the quasi-particles behaves as free particles, see Figure \ref{Fig1}.

Similarly to what discussed for $T(\eta_H)$, conjugation with $e^{A}$ does not substantially change the number of excitations. 
Indeed, the following lemma is proved in \cite[Sec. 5]{BBCS3}.
\begin{lemma} \label{prop:AellNgrow}
Suppose that $A$ is defined as in (\ref{eq:Aell1}). For any $k\in \bN$ there exists a constant $C >0$ such that the operator inequality  
\[ e^{-A} (\cN_++1)^k e^{A} \leq C (\cN_+ +1)^k   \]
holds true on $\cF_+^{\leq N}$, for all $\mu > \nu > 0$, and $N$ large enough. 
\end{lemma}

We use now the cubic phase $e^A$ to introduce a new excitation Hamiltonian, defining 
		\begin{equation*} 
		\cR_{N}:= e^{-A} \,\cG_{N}\,e^{A} \end{equation*}
on a dense subset of $\cF_+^{\leq N}$.  The definition of the excitation Hamiltonian $\cR_N$ corresponds to rewrite $N$-particle wave functions in the form 
\be \label{eq:trialcR}
\ps_N = U^* e^{A} \,T(\eta_H) \xi_N\,,
\ee with $\xi_N \in \cF^{\leq N}_+$. 
Conjugation with $e^A$ renormalizes the diagonal quadratic term and the cubic term on the r.h.s. of \eqref{eq:GNeff}, effectively replacing the singular potential $\widehat{V} (p/N)$ by a potential decaying already on momenta of order one.  The mechanism for this renormalization is similar to the one described around \eqref{eq:GNexp}. Again, expanding to second order we find
\begin{equation}\label{eq:mech} \cR_N =  e^{-A} \cG_N e^A \simeq \cG_N + [\cG_N , A] + \frac{1}{2} [[ \cG_N , A ] , A ] + \dots \, .  \end{equation}
From the canonical commutation relations (ignoring the fact that $A$ is cubic in generalized, rather than standard, field operators) we conclude that $[\cK, A]$ and $[\cV_N, A]$ are cubic and quintic in creation and annihilation operators, respectively. Some of the terms contributing to $[\cV_N, A]$ are not in normal order, i.e. they contain creation operators lying to the right of annihilation operators.  When we rearrange creation and annihilation operators to restore normal order, we generate an additional cubic contribution. There are therefore two cubic contributions arising from the first commutator $[\cG_N, A]$ on the r.h.s. of (\ref{eq:mech}). Moreover, the first commutator between the cubic term left in $\cG_N$ and $A$, and  the second commutator $[[\cH_N, A], A]$ produce the quadratic contributions that renormalizes the diagonal quadratic term in $\cG_N$. Indeed, one ends up with the following proposition, whose proof can be found in \cite[Sec. 8.6]{BBCS3}. \\

\begin{prop}  Let $V\in L^3(\bR^3)$ be compactly supported, pointwise non-negative and spherically symmetric. Then, for all choices of $\mu^{1/2} < \nu < \mu^{2/3}$, there exist  $\kappa, \a > 0$ and a constant $C>0$ such that  
\begin{equation} \begin{split} \label{eq:calR}   
{\cal R}_N =\; & {4 \pi \frak{a}_0 N} -  4 \pi \frak{a}_0 \frac{\cN_+^2}{N} + \cH_N \\
&+ { 8 \pi \frak{a}_0} \sum_{p \in \L^*_+,\, |p| \leq \mu} a^*_p a_p   \left(1 - \frac {\cN_+}{N} \right)  +  {4 \pi \frak{a}_0} \sum_{p \in \L^*_+,\, |p| \leq \mu} \big[b_p b_{-p} + b^*_p b^*_{-p}\big] \\
& + \frac{ 8 \pi \frak{a}_0}{\sqrt N} \sum_{ \substack{p, q \in \L^*_+\\ |p| \leq \mu,\, p \neq -q}}  \big[ b^*_{p+q} a^*_{-p} a_q  + \hc \big]   + \cE_{\cR_N}
\end{split}\end{equation}
with
\[\label{eq:RE} 
\pm \cE_{\cR_N} \leq  C\mu^{-\k} \,(\cH_N + 1) + C \mu^\a \,. \]
\end{prop}

We notice that $\cR_N$ is almost an excitation Hamiltonian for a mean field potential $8 \pi \frak{a}_0 \chi(|p| \leq \mu)$. More precisely, we define the function $\nu_\mu \in L^\infty (\Lambda)$ by setting 
\[ \nu_\mu(x) := 8\pi \frak{a}_0  \sum_{p \in \L^* :\, |p| \leq \mu} e^{i p \cdot x}\,. \]
In other words, $\nu_\mu$ is defined so that $\widehat{\nu}_\mu (p) = 8\pi \frak{a}_0$ for all $p \in \L^*$ with $|p| \leq \mu$ and $\widehat{\nu}_\mu (p) = 0$ otherwise. Observe, in particular, that $\widehat{\nu}_\mu (p) \geq 0$ for all $p \in \L^*$. Using \eqref{eq:U-rules} it is easy to check that most of the terms on the r.h.s. of \eqref{eq:calR} can be obtained by computing  $U N^{-1} \sum_{i<j}^N \nu_\mu(x_i - x_j) U^*$, and in fact we obtain the lower bound
\[ \begin{split}
\cR_N &  \geq     \frac 1 {N} \,  U \sum_{i<j}^N \nu_\mu(x_i - x_j) U^*  +  (1 - C \mu^{-\a} )  (\cH_N +1) \\
& \hskip 2cm - \frac{4 \pi \frak{a}_0}{N} \hskip -0.1cm\sum_{\substack{p,q, r \in \L^*_+ \\ |r| \leq \mu,\, r \neq -p, -q\\}}\hskip -0.2cm a^*_{p+r} a^*_q a_p a_{q+r} - C \, \frac{\cN_+^2}{N} - C \mu^\b\,.
\end{split}\]
Following a standard argument for mean field potentials with non negative Fourier transform (\eg \cite[Lemma 1]{Sei}) we find
\[
 \frac 1 N \sum_{i<j} \nu_\mu(x_i - x_j)  \geq  4 \pi \frak{a}_0 N  - C \mu^3\,.
\]
Using then the bound
\[ \begin{split}  \frac{4\pi \frak{a}_0}{N}  \sum_{\substack{p,q \in \L_+^*, |r| \leq \mu : \\ r \not = -p, -q}} \langle \xi , a_{p+r}^* a_q^* a_p a_{q+r} \xi \rangle &\leq \frac{C}{N} \sum_{\substack{p,q \in \L_+^* , |r|\leq \mu : \\ r \not = -p,-q}} \| a_{p+r} a_q \xi \| \| a_p a_{q+r} \xi \| \\ &\leq \frac{C \mu^3}{N} \| \cN_+ \xi \|^2\,, \end{split} \]
we conclude that  there exists $\b>0$ such that
\be \label{eq:RNfinal}
 \cR_N  \geq  4 \pi \frak{a}_0 N + \frac 12 \cH_N  \;  - \mu^3 \cN_+^2/N  - C \mu^\b\,.
\ee
If we were on a subspace of $\cF^{\leq N}_+$ with $\cN_+ \leq c N$, for sufficiently small $c>0$, we could conclude that 
\[
 \cR_N   \geq  4 \pi \frak{a}_0 N  + c \cN_+ - C
\]
this allowing us to show that $\cN_+$ is bounded on low energy states.  This observation suggests to apply localization techniques developed by Lewin-Nam-Serfaty-Solovej in \cite{LNSS} (inspired by previous work of Lieb-Solovej in \cite{LS}) based on localization of the number of excitations. On sectors with few excitations, we can control all the error terms in $\cR_N$ by the gap in the kinetic energy operator. On the other hand, on sectors with many excitations,  we are going to use that we do not have condensation,  and therefore the energy per particle  must be strictly larger than $4 \pi \frak{a}_0 N$ (due to the estimate  (\ref{eq:BEC1})), as described in the next section.

\section{Localization techniques and Bose-Einstein condensation}

Aim of this section is to explain how the application of localization techniques from \cite{LNSS} allows to show the optimal rate of condensation, and similar bounds for the energy of excitations.

Let $f,g: \bR \to [0;1]$ be smooth, with $f^2 (x) + g^2 (x)= 1$ for all $x \in \bR$. Moreover, assume that $f (x) = 0$ for $x > 1$ and $f (x) = 1$ for $x < 1/2$. We fix $M  = c N$  and we set $f_M = f (\cN_+ / M)$ and $g_M = g (\cN_+ / M)$. It follows from \cite[Proposition 4.3]{BBCS3}  that
\begin{equation}\label{eq:cGN-1} 
\begin{split}\cG_{N} - 4 \pi \frak{a}_0 N  & \geq f_M (\cG_{N} - 4 \pi \frak{a}_0 N) f_M + g_M (\cG_{N} - 4 \pi \frak{a}_0 N) g_M \\
&\hskip 6cm  - \frac{C \m^{1/2}}{N^{2}}  (\cH_N + 1) \end{split}\end{equation}
for $\mu$, $N \in \bN $ and $M \in \bN$ large enough. 
To bound  $f_M  \cG_{N} f_M$ we conjugate $\cG_N$ by $e^{-A}$ and use the lower bound \eqref{eq:RNfinal}
\be \label{eq:fMGN}
\begin{split}
f_M  \cG_{N} f_M & \geq   f_M  e^A \, \cR_{N} \, e^{-A} f_M   \\
& \geq  4 \pi \frak{a}_0 N  f^2_M +   f_M e^A \left[ \frac 1 2 \cH_N - \mu^3 \cN_+^2/N - C \mu^\a \right] e^{-A} f_M \\
& \geq   4 \pi \frak{a}_0 N  f^2_M +   f_M e^A \left[ \frac 1 2 \cH_N - \mu^\k \cN_+  \right] e^{-A} f_M  -  C \mu^\a f_M^2
\end{split}\ee
where we used Lemma \ref{prop:AellNgrow} and chose $M=\mu^{-3-\k} N$. Using the gap in the kinetic energy and once more  Lemma  \ref{prop:AellNgrow} we conclude that for $\mu$ large enough
\be \label{eq:fMGN}
f_M  \cG_{N} f_M  \geq  4 \pi \frak{a}_0 N  f^2_M  + C f^2_M \cN_+ - C \mu^\a f_M^2\,.
\ee
On $g_M$ using (\ref{eq:BEC1}) one can claim that there exists a constant $C>0$ such that
\be  \label{eq:gM-bd}
g_M  \cG_{N} g_M \geq   4 \pi \frak{a}_0 N  g^2_M + C g^2_M N 
\ee
for all $N$ sufficiently large. Indeed, if this was not the case one could build, starting from an excitation vector $ \xi \in \cF^{\leq N}_{\geq M/2}$ with at least $M/2= \mu^{-3-\k} N /2$ particles, an approximate ground state of $H_N$.  But this would contradict (\ref{eq:BEC1}) since the ratio between the expected number of excitations on $\xi$ and the total number of particles would not go to zero as $N \to \io$. We refer the reader to \cite[Sect. 6]{BBCS3} for details. From \eqref{eq:gM-bd}, using $\cN_+ \leq N$ we get
\be  \label{eq:gM-bd2}
g_M  \cG_{N} g_M \geq   4 \pi \frak{a}_0 N  g^2_M + C \cN_+ g^2_M  \,.
\ee
Inserting (\ref{eq:fMGN}) and (\ref{eq:gM-bd2}) on the r.h.s. of (\ref{eq:cGN-1}), we obtain that 
\begin{equation}\label{eq:lbd} \cG_{N}  \geq  4\pi \frak{a}_0 N + C \cN_+  - C N^{-2} \cH_N - C \end{equation}
for $N$ large enough (the constants $C$ are now allowed to depend on $\mu$ and $\nu$, since the cutoff has been fixed once and for always after (\ref{eq:fMGN})). Interpolating (\ref{eq:lbd}) with the lower bound
\[
\cG_N \geq  4\pi \frak{a}_0 N  + \frac 12 \cH_N - C \cN_+ - C\,,
\]
obtained using \eqref{eq:Gbd0},  we get 
\begin{equation}\label{eq:cGN-fin} 
\cG_{N}  \geq 4\pi \frak{a}_0 N+ c\, \cN_+ - C \,. \end{equation}
The condensation bound follows easily from \eqref{eq:cGN-fin}. Let now $\psi_N \in L^2_s (\Lambda^N)$ with $\| \psi_N \| =1$ and
\[ \langle \psi_N , H_N \psi_N \rangle \leq 4 \pi \frak{a}_0 N + \zeta\,. \]
Recalling that  $\cG_{N} = e^{-B(\eta_H)}  U_N H_N U_N^* e^{B(\eta_H)}$ and defining 
the excitation vector $\xi_N = e^{-B(\eta_H)} U_N \psi_N$, we have 
\be \label{eq:Nplus}
 \langle \xi_N, \cN_+ \xi_N \rangle \leq C \langle \xi_N, (\cG_{N} - 4\pi \frak{a}_0 N) \xi_N \rangle + C \leq C  (1 + \zeta)\,. \ee
Following a strategy similar to one described for the mean-field case in Sec. \ref{sec:MF} we can show the following stronger bounds on excitation vectors, see \cite[Sec. 4]{BBCS4} for their proof.  
\begin{prop}\label{prop:hpN}
Let $V \in L^3 (\bR^3)$ be non-negative, compactly supported and spherically symmetric. Let $E_N$ be the ground state energy of the Hamiltonian $H_N$ defined in (\ref{eq:Hmom}) (or, equivalently, in (\ref{eq:Ham0})). Let $\psi_N \in L^2_s (\Lambda^N)$ with $\| \psi_N \| = 1$ belong to the spectral subspace of $H_N$ with energies below $E_N + \zeta$, for some $\zeta > 0$, i.e. 
\begin{equation}\label{eq:psi-space} \psi_N = {\bf 1}_{(-\infty ; E_N + \zeta]} (H_N) \psi_N \, . 
\end{equation}
Let $\xi_N = e^{-B(\eta)} U_N \psi_N$ be the renormalized excitation vector associated with $\psi_N$. Then, for any $k\in\mathbb N$ there exists a constant $C > 0$ such that 
\[ \label{eq:hpN} \langle \xi_N, (\cN_+ +1)^k (\cH_N+1) \xi_N \rangle \leq C (1 + \zeta^{k+1}) \, . \]
\end{prop}
Using these bounds we are now in the position to establish the validity of Bogolibov theory in the Gross-Pitaevskii regime, as pictured in the next section. Notice also that the bound \eqref{eq:Nplus} also implies an improved bound for the trace norm convergence in \eqref{eq:BEC1}. In fact, if $\gamma_N$ denotes the one-particle reduced density matrix associated with $\psi_N$, we obtain 
\[ \begin{split}  \label{eq:BEC}
1 - \langle \ph_0, \gamma_N \ph_0 \rangle &= 1 - \frac{1}{N} \langle \psi_N, a^* (\ph_0) a (\ph_0) \psi_N \rangle \\ &= 1 - \frac{1}{N} \langle U_N^* e^{B(\eta_H)} \xi_N, a^* (\ph_0) a(\ph_0) U_N^* e^{B(\eta_H)} \xi_N \rangle \\ &= \frac{1}{N} \langle e^{B(\eta_H)} \xi_N, \cN_+ e^{B(\eta_H)} \xi_N \rangle \\
& \leq \frac{C}{N} \langle \xi_N , \cN_+ \xi_N \rangle \leq \frac{C(K+1)}{N} \end{split} \]
where in the last line we used Lemma \ref{lm:Ngrow}. 

\section{Bogoliubov theory}

In this section we sketch the proof of Theorem \ref{thm:main}, as proved in \cite{BBCS4}; we refer  to the review \cite{Schlein} for a more extended presentation of this part, which is only slightly modified whenever we remove the assumption of smallness of the potential. The key idea  is that using the bounds in Prop.\ref{prop:hpN}, we can give a second look to the excitation Hamiltonian $\cG_{N} = e^{-B(\eta_H)}  U_N H_N U_N^* e^{B(\eta_H)}$, and identify terms which go to zero as $N\to \io$ on low energy states. More precisely, one finds
\be \label{eq:GNsecond}
\cG_N = C_N + \cQ_N + \cC_N+ \cV_N + \d_N
\ee
where $C_N$ is a constant, $\cQ_N$ is quadratic,
\[
\cC_N = \frac{1}{\sqrt N} \sum_{p, q \in \L^*_+, p\neq -q} \widehat{V}(p/N) \,b^*_{p+q}b^*_{-p} \big( \g_q b_q + \s_q b^*_{-q}\big) + \hc 
\]
with $\g_p = \cosh(\eta_H((p))$ and $\s_p = \sinh(\eta_H(p))$, and the error term satisfies the bound
\[
\pm \d_N \leq \frac C {\sqrt N} \Big[ (\cH_N + 1)(\cN_++1) + (\cN_++1)^3 \Big]\,.
\]
If there was no cubic term on the r.h.s. of \eqref{eq:GNsecond} we could obtain the ground state energy and spectrum of $\cG_N$ just by diagonalizing a quadratic Hamiltonian. In fact the quartic interaction can be bounded from above by $C N^{-1} $on suitable states (see \cite[Lemma 6.1]{BBCS4}), and can be neglected due to positivity of the interaction as lower bounds are concerned. 

Once more, the strategy to renormalize the large (order one) cubic term is to conjugate $\cG_N$ by a suitable unitary operator, given by the exponential of the operator 
\[
\tl A = \frac{1}{\sqrt N} \sum_{\substack{p, q \in \L^*_+ \\ |r| \geq \sqrt N,\, |v| < \sqrt{N} }} \eta_r \Big[ \s_v b^*_{r+v}b^*_{-r} \big( \g_v b_v + \s_v b^*_{-v}\big) - \hc \Big]\,.
\]
Notice that, as in the definition of $A$ in \eqref{eq:Aell1}, the operator $\tl A$ describes the scattering between two excitations with high momenta, one excitation with low momenta and one particle in the condensate, but with a different notion of ``high'' and   ``small'' momenta with respect to \eqref{eq:Aell1}. We introduce a new excitation Hamiltonian
\be \label{eq:JN}
\cJ_N = e^{-\tl A} \cG_N e^{\tl A} =e^{-\tl A}  T^*(\eta_H) U_N H_N U_N^* T(\eta_H) e^{\tl A}  : \quad  \cF^{\leq N}_+ \to\cF^{\leq N}_+\,.
\ee
The latter can be decomposed as
\[
\cJ_N = \tl C_N + \tl \cQ_N + \cV_N + \tl \d_N\,,
\]
where $\tl C_N $ and $\tl \cQ_N $ are constant and quadratic in annihilation and creation operators, and where 
\[
\pm \tl \d_N \leq  C N^{-1/4} \Big[ (\cH_N + 1)(\cN_++1) + (\cN_++1)^3 \Big]\,.
\]
In particular the expression of the quadratic part makes evident the effect of the renormalization obtained by conjugating  $\cL_N$ with the unitary operators $T(\eta)$ and $e^{-\tl A}$. We have in fact: 
\[
\tl \cQ_N  = \sum_{p \in \L^*_+} \Big[  F_p b^*_p b_p + G_p \big( b^*_{p} b^*_{-p} + b_p b_{-p} \big) \Big]
\]
with
\[ \begin{split}
F_p & = p^2 \big(\g^2_p + \s^2_p \big) + \big( \widehat V(\cdot/N) \ast \widehat f_N \big)_p \big( \g_p + \s_p \big)^2 \\
G_p &= 2 p^2 \g_p \s_p + \big( \widehat V(\cdot/N) \ast \widehat f_N \big)_p \big( \g_p + \s_p \big)^2\,.
\end{split}\]
We see that the Fourier transform of the interaction potential $\widehat V(p/N)$ has been replaced everywhere by $ \big( \widehat V(\cdot/N) \ast \widehat f_N \big)_p$, whose value for $p=0$ is related to the scattering length $\frak{a}_0$ through the relation \eqref{eq:intVf}. One can check that  for all $p \in \L^*_+$
\[
p^2/2 \leq F_p \leq  C (1+p^2) \,, \qquad |G_p| \leq C/p^2 \,, \quad |G_p| < F_p \,,
\]
and therefore we can introduce coefficients $\t_p \in \bR$ such that
\[ \label{def:taup}
\tanh(2\t_p) = - \frac{G_p}{F_p}
\]
for all $p \in \L^*_+$.  Using these coefficients, we define the generalized Bogoliubov transformation $e^{B(\tau)}: \cF_+^{\leq N}\to \cF_+^{\leq N}$ with
        \begin{equation*}
        T(\tau):= \exp \left[\frac{1}{2}\sum_{p\in\Lambda^*_+}\tau_{p}\big(b^*_{-p}b^*_p-b_{-p}b_p\big) \right] \, . 
 \end{equation*}
Notice that, since $|\t_p| \leq C |p|^{-4}$ for all $p \in \L^*_+$ we can show that (see \cite[Lemma 5.2]{BBCS4})
\begin{equation*}\label{eq:err-con} T^*(\tau) (\cN_+ +1)(\cH_N+1) T(\tau) \leq C  (\cN_+ +1)(\cH_N+1)\,, \end{equation*}
that is the generalized Bogoliubov transformation $T(\t)$ does not change substantially neither the number nor the energy of the excitations. 
Conjugation of  the excitation Hamiltonian $\cJ_N$ defined in (\ref{eq:JN}) with 
$T(\tau)$ leads to the excitation Hamiltonian
\[
\cM_N =  T^*(\t) \cJ_N T(\t)=T^*(\t) e^{-\tl A}\, T^*(\eta_H) U_N H_N U_N^* T(\eta_H) e^{\tl A}\, T(\t) : \quad  \cF^{\leq N}_+ \to\cF^{\leq N}_+ \,. 
\]
One finally finds
\[ \begin{split}
\cM_N =\, & 4 \pi \frak{a}_0 (N-1) + e_\L \frak{a}_0^2 \\
& +  \frac12\sum_{p\in\Lambda_+^*}\bigg[ \sqrt{p^4+16\pi \frak{a}_0   p^2} -p^2-8\pi \frak{a}_0  + \frac{(8\pi \frak{a}_0 )^2}{2p^2}   \bigg] \\
& + \sum_{p\in\Lambda_+^*} \sqrt{p^4 + 16 \pi \frak{a}_0 p^2} a^*_p a_p + \cV_N + \d'_N
\end{split}\]
with
\[
\pm \d'_N \leq C N^{-1/4} \Big[ (\cH_N + 1)(\cN_++1) + (\cN_++1)^3 \Big]\,.
\]
Theorem \ref{thm:main} follows from min-max principle, since on low energy states of the diagonal Hamiltonian, we find $\cV_N \leq C N^{-1} (\zeta + 1)^{7/2}$(with $\zeta$ entering in the spectral assumption \eqref{eq:psi-space}).

The results of Theorem \ref{thm:main}, together with standard arguments as in \cite[Section 7]{GS}, also provide an approximation for the eigenvectors corresponding to low energy states. In particular,  if $\psi_N$ denotes a ground state vector of the Hamiltonian $H_N$, one can show that there exists a phase $\omega \in [0;2\pi)$ such that 
\[\label{eq:gs-appro} 
\big\| \psi_N - e^{i\omega} U_N^* \,T(\eta_H)\, e^{\tl A}\, T(\t) \Omega \big\|^2 \leq \frac{C}{\theta_1 - \theta_0} N^{-1/4}\, ,  \]
where  $\theta_0 \leq \theta_1 \leq \dots $ denote the ordered eigenvalues of $H_N$.  

It is interesting to compare \eqref{eq:gs-appro} with the approximation for the ground state vector within Bogoliubov approximation, that would have been of the form $U^*_N \tl T(\tl \tau) \O$, with
\[ \label{eq:deftlT}
\tl T(\eta_H) = \exp \left[ \frac 12 \sum_{|p|\geq \mu}  \tl \t_p \big( a^*_p a^*_{-p} - a_p a_{-p})\right]
\]
a usual Bogoliubov transformation, and coefficients $\tl \tau_p \in \bR$ that in the Gross-Pitaevskii regime would be defined by
\[
\tanh (2 \tl \tau_p) = - \frac{\widehat V(p/N)}{p^2 + \widehat V(p/N)}\,,
\]
see \cite[Appendix A]{LSSY}. The unitary transformation $\tl T(\tl \tau)$ is the one that diagonalizes the quadratic terms in Bogoliubov Hamiltonian, so it has the same role of the transformation $T(\tau)$ in our approach.  On the other side, while the kernel $\tl \tau_p$ has a large $H^1$- norm (similarly to the kernel $\eta_p$ defined in \eqref{eq:defeta}), the kernel $\tau_p$ has both the $L^2$ and $H^1$ norms uniformly bounded in $N$. Hence, the diagonalizing unitary transformation $T(\tau)$ does not change substantially neither the number nor the energy of excitations.
We see that the trick to take into account for the correlations among excitations neglected in Bogoliubov theory was to implement two additional unitary transformations $T(\eta_H)$ and $e^{\tl A}$ which have extracted the large energy contained in the cubic and quartic terms. In particular, one of the consequences of the action of  $T(\eta_H)$ and $e^{\tl A}$  is the renormalization of the quadratic terms of the excitation Hamiltonian $\cR_N$, leading to the appearance of the convolution  $\big( \widehat V(\cdot/N) \ast \widehat f_N \big)_p$ in the definition of the coefficients $\t_p$.


\end{document}